\documentstyle[12pt]{article}

\newcommand{\pagenumber}{\pagestyle{plain}\setcounter{page}{1}} 

\def\fnote#1#2{\begingroup\def\thefootnote{#1}\footnote{#2}
    \addtocounter{footnote}{-1}\endgroup}
\def\sppt{makoto@sbitp.ucsb.edu}

\begin{document}

\pagestyle{empty}

\begin{flushright}
    NSF-ITP-94-66\\
    hep-th/9406079\\
\end{flushright}
\vspace{24pt}
\begin{center}
{\bf Higher Order Correction to the GHS String Black Hole}

\vspace{36pt}
Makoto Natsuume\fnote{*}{\sppt}

\vspace{6pt}
{\sl Institute for Theoretical Physics \\
University of California \\
Santa Barbara, California 93106-4030 }
\vspace{48pt}

\underline{ABSTRACT}

\end{center}

\vspace{24pt}
We study the order $\alpha'$ correction to the string black hole
found by Garfinkle, Horowitz, and Strominger.
We include all operators of dimension up to four in the Lagrangian,
and use the field redefinition technique which facilitates the analysis.
A mass correction, which is implied by the work of
Giddings, Polchinski, and Strominger, is found
for the extremal GHS black hole.

\vfill

\newpage
\pagenumber                                         
\baselineskip=20pt 

In recent years,
the string black hole solution found by
Garfinkle {\it et al.} (GHS solution hereafter)
has attracted much attention \cite{ghs,gibbons,horowitz}.
They solved the leading $\alpha'$ action
of the low-energy effective theory for the heterotic string.
The solution describes
a four dimensional magnetically charged black hole
coupled with the dilaton.

On the other hand, Giddings {\it et al.} \cite{gps}
has obtained the exact solution for the black hole in the extremal limit
(GPS solution hereafter).
GPS construct an exact CFT
(so, the solution includes the effects of all higher order
$\alpha'$ terms) which describes the throat region
of the extremal GHS black hole.
One new feature of the GPS solution compared with the GHS solution
is the existence of a neutral throat.
The neutral throat is possible for the GPS solution
since $R=|Q^{2}-1|^{\frac{1}{2}}$,
where $R$ is the throat radius and $Q$ is the monopole charge.
However, the GHS solution does not allow the neutral throat
since $R=Q(=2M)$.
The GPS solution is a solution of the full effective theory,
but the GHS solution is only a leading order solution;
so any difference between these two solutions
should come from the higher order terms in the effective theory.
Thus, the GPS solution suggests
that there exists correction to the black hole mass due to the higher
order $\alpha'$ terms to make the neutral throat possible.
Unfortunately, the GPS solution does not connect to
the asymptotic region;
so, they cannot get the mass correction.
The purpose of this paper is to obtain the mass correction
by a different approach.

To obtain the correction,
we shall keep $O(\alpha')$ terms as well in the effective action
and perform $\alpha'$ perturbative expansion around the GHS background.
Without the detailed knowledge of the effective action, we will show that the
extremal GHS black hole gets a correction in mass
given by $M = Q/2 - \alpha'/40Q$.

In order to show this result, we will thoroughly employ
the field redefinition technique for analysis.
This is a useful technique to simplify higher order effective actions.
Even though the field redefinitions change the action
and fields like the metric,
it does not change the physics;
for instance, the $S$-matrix is invariant under the field redefinitions
by the equivalence theorem \cite{itzykson,tseytlin}.
Therefore, when one works with higher orders in the effective theory,
one can simplify the effective theory by transforming the original
action into a simpler one by the field redefinitions.
Although this technique is illustrated for the magnetically charged
black hole, the technique itself is general
and is useful for the other ``dirty" black holes.

In the extremal limit, the GHS solution is given by \cite{ghs}
\begin{eqnarray}
d^{2}s_{string} & = &
	- dt^{2} + \left(1-\frac{Q}{r}\right)^{-2} dr^{2}
	+ r^{2} d\Omega
		\nonumber \\
e^{-2\phi} & = & 1-\frac{Q}{r}
		\label{eq:ghs_string}
		\\
F & = & Q \sin\theta d\theta \wedge d\varphi
		\nonumber
\end{eqnarray}
in the string metric, and
\begin{equation}
d^{2}s =
	- \left(1-\frac{Q}{r}\right) dt^{2}
	+ \left(1-\frac{Q}{r}\right)^{-1} dr^{2}
        + r^{2}\left(1-\frac{Q}{r}\right) d\Omega
\end{equation}
in the Einstein metric.
Here, the string metric $g_{\mu \nu}$ and
the Einstein metric $\tilde{g}_{\mu \nu}$ are related by
$g_{\mu \nu} = e^{2\phi} \tilde{g}_{\mu \nu}$.
$\phi_{0}$, the asymptotic value of the dilaton,
is set to zero for simplicity.
The mass of the black hole is given by $M=Q/2$.

Our starting point is the most general action to the order $\alpha'$
with all possible independent terms:
\begin{equation}
S = \int d^{4}x \sqrt{-g} e^{-2\phi}
  \left\{ {\cal L}_{2} + \sqrt{\alpha'} {\cal L}_{3} + \alpha' {\cal L}_{4}
	\right\},
\label{eq:full_action}
\end{equation}
where ${\cal L}_{i}$ denotes the contributions of $i$ derivative operators.
${\cal L}_{2}$ is the leading order Lagrangian given by
\begin{equation}
{\cal L}_{2} = R + 4(\nabla\phi)^{2} - \frac{1}{2} F^{2}.
\end{equation}
We need to include all operators
which contain at most four derivatives
because the effective theory expansion is simply a derivative expansion for the
above choice of the field normalizations \cite{polchinski}. The field
normalization of $A_{\mu}$ indicates that the gauge field has zero mass
dimension. This is different from the conventional normalization \cite{gross}
where $A_{\mu}$ has one mass dimension (thus, $F^{2}$ is an $O(\alpha')$ term).
In other words, we implicitly made the $\alpha'$-rescaling of $A_{\mu}$, which
implies that the charge $Q$ is not small, {\it i.e.\/}, $Q$ is $O(1)$ instead
of $O(\sqrt{\alpha'})$. The rescaling is natural since we consider a large mass
black hole for the $\alpha'$ perturbation to be valid.

${\cal L}_{3}$ and ${\cal L}_{4}$ are given by
\begin{eqnarray}
{\cal L}_{3} & = & a_{1} F^{\mu \nu}F_{\nu \rho}F^{\rho}_{~~\mu},
			\\
{\cal L}_{4} & = &
a_{2} R^{\mu \nu \rho \sigma}R_{\mu \nu \rho \sigma}
+ a_{3} R^{\mu \nu}R_{\mu \nu}
+ a_{4} R^{2}
	\nonumber \\
	& & \mbox{}
+ a_{5} R^{\mu \nu}\nabla_{\mu}\phi\nabla_{\nu}\phi
+ a_{6} R(\nabla\phi)^{2}
+ a_{7} R\nabla^{2}\phi
	\nonumber \\
	& & \mbox{}
+ a_{8} (\nabla^{2}\phi)^{2}
+ a_{9} (\nabla\phi)^{2} \nabla^{2}\phi
+ a_{10} (\nabla\phi)^{4}
	\nonumber\\
	& & \mbox{}
+ a_{11} (\nabla_{\mu}F^{\mu \rho})(\nabla^{\nu}F_{\nu \rho})
+ a_{12} F^{\mu \nu}F_{\nu \rho}F^{\rho \sigma}F_{\sigma \mu}
+ a_{13} (F^{\mu \nu}F_{\mu \nu})^{2}
	\nonumber\\
	& & \mbox{}
+ a_{14} R^{\mu \nu \rho \sigma}F_{\mu \nu}F_{\rho \sigma}
+ a_{15} R^{\mu \nu}F_{\mu \rho}F_{\nu}^{~~\rho}
+ a_{16} RF^{2}
        \nonumber \\
        & & \mbox{}
+ a_{17} F^{2}(\nabla\phi)^{2}
+ a_{18} F^{2}(\nabla^{2}\phi)
+ a_{19} \nabla_{\mu}\phi\nabla_{\nu}\phi F^{\mu \rho} F^{\nu}_{~~\rho}.
\end{eqnarray}
Using the leading order GHS solution,
one can easily check that every operator has the same order
of magnitude.

We did not include terms which are proportional to
the three-form field strength $H_{\mu\nu\rho}$, where
\begin{equation}
H_{\mu\nu\rho} = \partial_{\,[\rho}B_{\mu\nu]}
	+\frac{1}{4}(\Omega^{(L)}_{\mu\nu\rho}-\Omega^{(Y)}_{\mu\nu\rho}).
\end{equation}
$\Omega^{(L)}$ and $\Omega^{(Y)}$ are the Lorentz and Yang-Mills
Chern-Simons forms respectively.
Since the Chern-Simons forms linearly couple with $B_{\mu\nu}$,
the Chern-Simons forms act as sources for $H_{\mu\nu\rho}$;
therefore, we may not be able to simply set $H=0$.
However, for the spherically symmetric metrics we examine below,
the Lorentz Chern-Simons form can be expressed
as the exterior derivative of a three-form;
thus we can absorb it into the definition of $B_{\mu\nu}$
\cite{campbell2}.
Also, the Yang-Mills Chern-Simons form vanishes for the purely magnetic case.
For these reasons, it is consistent to set $H_{\mu\nu\rho}$ to zero.

We assume spherical symmetry; the Bianchi identity then implies that
$F_{\mu \nu}$ is unchanged.
Thus, the $a_{1}, a_{11},$ and $a_{19}$ terms vanish;
also, the $a_{12}$ term is no longer independent of the $a_{13}$ term.
We will not consider these terms further,
so the above conditions leave fifteen $O(\alpha')$ terms
($a_{2},\ldots, a_{10}$ and $a_{13},\ldots, a_{18}$).

The coefficients of higher order terms
are free parameters in general due to field redefinition ambiguity.
To see this, consider the most general field redefinitions:
\begin{eqnarray}
g_{\mu \nu} & = & g'_{\mu \nu} + \alpha' T_{\mu \nu}(g', \phi', F')
		+ O(\alpha'^{2})
	\nonumber \\
      \phi  & = & \phi + \alpha' T(g', \phi', F')
		+ O(\alpha'^{2})
	\label{eq:field_redef1}
	\\
      A_{\mu} & = & A'_{\mu} + O(\alpha'^{2}),
	\nonumber
\end{eqnarray}
where
\begin{eqnarray}
T_{\mu \nu}
  & = & g_{1}R_{\mu \nu} + g_{2}\nabla_{\mu}\phi\nabla_{\nu}\phi
			 + g_{3}F_{\mu \rho}F_{\nu}^{~~\rho}
	\nonumber \\
  &   & \mbox{} + g_{\mu \nu}
	\left\{
		g_{4}R + g_{5}\nabla^{2}\phi + g_{6}(\nabla\phi)^{2}
		       + g_{7}F^{2}
	\right\}
	\label{eq:field_redef2}
	\\
T & = & d_{1}R + d_{2}\nabla^{2}\phi + d_{3}(\nabla\phi)^{2} + d_{4}F^{2}.
	\nonumber
\end{eqnarray}
Here, $g_{i}$ and $d_{i}$ are free parameters.
We have used spherical symmetry
to eliminate possible field redefinition for $A_{\mu}$ at $O(\alpha')$.
Substituting (\ref{eq:field_redef1}) and (\ref{eq:field_redef2})
into (\ref{eq:full_action}),
one finds that ${\cal L}_{4}$ retains
its form but the coefficients $a_{i}$ change in general.
The explicit result can be found in appendix.
\footnote{The coefficient changes for the gravity-dilaton part
have been considered by Metsaev and Tseytlin \cite{tseytlin2}.}

In the literature, one often claims that the ambiguity is resolved
by choosing the Gauss-Bonnet scheme for curvature squared terms,
{\it i.e.\/}, by taking $a_{2}=-1/2$ and $a_{3}=1/8$.
The justification is the argument by Zwiebach \cite{zwiebach};
he argued that only the Gauss-Bonnet combination
gives a ghost-free theory in the weak field expansion.
The argument is actually irrelevant \cite{gross}
since our effective action is a perturbative expansion
in powers of momentum;
the perturbation itself is not valid at the energy of the apparent ghost.

While we are not able to resolve the ambiguity,
this ambiguity does not matter
as long as physical quantities measured at infinity are concerned.
This is because the field redefinitions do not alter these quantities.
In this sense,
the only meaningful quantities to evaluate
in higher order effective theories are
relations of physical quantities,
like mass-charge, mass-temperature relations, and so on.
Moreover, because physical results are unchanged under the redefinitions,
what one should do is to find the simplest Lagrangian one can reach by
the redefinitions to simplify calculations.

As a check of the above statement,
we show that mass and charge are unchanged
under the field redefinitions (\ref{eq:field_redef1}).
The monopole charge is of course invariant
since there is no possible field redefinition for $A_{\mu}$
at this order.
The gravitational mass $M_{G}$ and the inertial mass $M_{I}$ are
defined by the asymptotic behavior of the Einstein metric \cite{mtw};
\begin{equation}
\tilde{g}_{00} = -1 + \frac{2M_{G}}{r} + O(r^{-2}),\;\;\;
\tilde{g}_{11} =  1 + \frac{2M_{I}}{r} + O(r^{-2}).
\label{eq:mass}
\end{equation}
In the string metric, $M_{G}$ depends on $O(r^{-1})$ terms
in both $g_{00}$ and $\phi$
(and similarly for $M_{I}$).
Using the GHS solution (\ref{eq:ghs_string}), one can check
that the field redefinitions affect the terms of
order $r^{-3}$ or higher for both the metric and the dilaton;
therefore, black hole mass is invariant under the field redefinitions.

We now simplify ${\cal L}_{4}$ using the field redefinitions. There are
originally fifteen terms in ${\cal L}_{4}$. From the explicit calculation,
$a_{2}$ and $a_{14}$ are invariant under the redefinitions. The coefficients of
these terms are thus determined from a standard $S$-matrix calculation; for the
heterotic string, $a_{2}=1/8$ and $a_{14}=0$ \cite{gross}. This leaves thirteen
field redefinition dependent terms (ambiguous terms). The field redefinitions
have eleven free parameters, so one might expect that
it is possible to remove all the ambiguous terms except two
by appropriate field redefinitions. This conclusion is in fact correct, but the
reasoning is wrong. There is a subtlety in the counting because variations of
$a_{i}$ are not independent of each other (see appendix).

First, consider the gravity-dilaton part of ${\cal L}_{4}$. There are eight
ambiguous terms ($a_{3},\ldots, a_{10}$) and the field redefinitions have eight
parameters as well. ($F$-dependent redefinitions do not affect the
gravity-dilaton action.) However, the variations of these $a_{i}$ satisfy
(\ref{eq:invariant1}); one $\delta a_{i}$ is fixed completely once the rest are
chosen. Therefore, one term cannot be eliminated in the gravity-dilaton action.
Fortunately, Metsaev and Tseytlin show that the remaining term vanishes after
the field redefinitions \cite{tseytlin2}; thus, all ambiguous terms are
removed. But it is actually useful to work with the Gauss-Bonnet scheme instead
of keeping only the $a_{2}$ term. This scheme is useful because field equations
are at most second order in derivatives, which is reminiscent of the claim that
the scheme gives the ``ghost free" theory.

Next, consider the terms coupled with the gauge field. There are five ambiguous
terms in the action ($a_{13}, a_{15},\ldots, a_{18}$) and three $F$-dependent
field redefinition parameters ($g_{3}, g_{7}$, and $d_{4}$).\footnote{In the
above discussion of the gravity-dilaton action, one field redefinition
parameter is not used since we fix only seven $a_{i}$ in the action. However,
the dependence on this parameter disappears after the $F$-independent
redefinitions.} This leaves two ambiguous terms, but another relation
(\ref{eq:invariant2}) partly determines which two $a_{i}$ should be left. Since
the variations of $a_{13}, a_{15}, a_{17}$, and $a_{18}$ contribute to
(\ref{eq:invariant2}), at least one of them should be left. We will keep the
$a_{13}$ and $a_{17}$ terms since the equations of motions become simple.

Consequently, we have reached the following simple Lagrangian:
\begin{equation}
{\cal L}_{4} =
	\frac{1}{8}(R^{\mu \nu \rho \sigma}R_{\mu \nu \rho \sigma}
		- 4R^{\mu \nu}R_{\mu \nu} + R^{2})
		+ b (F^{2})^{2} + c F^{2}(\nabla\phi)^{2}.
\label{eq:GB_scheme}
\end{equation}
$b$ and $c$ are the parameters that we can fix by an $S$-matrix calculation
\cite{gross} and (\ref{eq:bc}). However, this step is unnecessary. The
mass-charge relation we are interested in will not depend on $b$ after a
coordinate transformation and $c$ will be fixed once we impose that the
solution behaves like the GPS solution in the throat region.

We solve field equations by perturbing around the GHS background.
Take the Schwarzschild gauge in the string metric
\begin{equation}
d^{2}s_{string}
	=  - e^{2\Phi} dt^{2} + e^{2\Lambda} dr^{2} + r^{2} d\Omega
\end{equation}
and expand the metric functions in $\hat{\alpha} \equiv \alpha'/Q^{2}$:
\begin{eqnarray}
\Phi & = & \hat{\alpha}\Phi_{2} + \cdots
	\nonumber \\
\Lambda & = & \Lambda_{1} + \hat{\alpha}\Lambda_{2} + \cdots
        \nonumber \\
\phi & = & \phi_{1} + \frac{\hat{\alpha}}{2}(\Phi_{2}-\phi_{2}) + \cdots,
	\nonumber
\end{eqnarray}
where $\Lambda_{1}$ and $\phi_{1}$ are given by the GHS solution
(Note $\Phi_{1}=0$).
The somewhat artificial choice of $\phi_{2}$ is useful to simplify
field equations.
Then, the Lagrangian (\ref{eq:GB_scheme}) becomes
\begin{eqnarray}
{\cal L} & \propto &
	- f^{2}\Phi_{2}'^{2}
	- x^{3}f \left\{ 2-(1+2c)x \right\}\Phi_{2}'
		\nonumber \\
	& & \mbox{} + f^{2}\phi_{2}'^{2}
	+ \frac{1}{5} x^{4} \left\{ -10c + (9c-b)x \right\}\phi_{2}'
	- 2f \left( 1-\frac{2}{x} \right) \phi_{2}'\Lambda_{2}
		\nonumber \\
	& & \mbox{} - \left( 1-\frac{2}{x^{2}} \right)\Lambda_{2}^{2}
	+ (b-c)x^{4}\Lambda_{2},
\end{eqnarray}
where $f = 1-Q/r$, $x = Q/r$, and the prime denotes a derivative with respect
to $x$.

The only solution regular at the horizon $r=Q$ is given by
\begin{eqnarray}
\Phi_{2} & = & - \frac{1+2c}{8}x^{4}
	+\frac{1-2c}{2}\left\{ \frac{1}{3}x^{3}+\frac{1}{2}x^{2}+x
	+\ln(1-x) \right\}
	\\
\Lambda_{2} & = & - \frac{b}{20}\frac{x}{1-x}
	(6x^{4}+5x^{3}+4x^{2}+3x+2)
\nonumber \\
	& & \mbox{}\hspace{1.5in} - \frac{c}{20} x(14x^{3}+9x^{2}+5x+2)
	\\
\phi_{2} & = & \frac{b\,x}{1-x} + 2(b-c)\ln(1-x)
\nonumber \\
	& & \mbox{} + \frac{11b-19c}{10} x + \frac{b-4c}{5} x^{2}
	- \frac{b+21c}{60} x^{3} - \frac{b+c}{20} x^{4}.
\end{eqnarray}
Even though $1/(1-x)$ terms in $\Lambda_{2}$ and $\phi_{2}$
look like singular perturbations,
they are not since a coordinate transformation
$x \rightarrow x - b\,\hat{\alpha}x( 6x^{4}+5x^{3}+4x^{2}+3x+2 )/20$
removes the terms.
Also, the above solution suggests $c=1/2$;
otherwise, $\Phi_{2}$ contains a term proportional to $\ln(1-x)$.
Such a term should be absent in the light of the GPS solution.
GPS claim that the extremal solution in the throat region
is a product of an $rt$ CFT (the linear dilaton theory)
and an angular CFT.
In particular, $g_{00} \rightarrow -1$.
If the $\ln(1-x)$ term existed in $\Phi_{2}$, our solution would not
give the linear dilaton theory in the throat region
under any choice of field redefinition,
{\it i.e.\/}, $g_{00}$ would not approach to $-1$.
On the other hand, $\ln(1-x)$ in $\phi_{2}$ is safe;
it just shifts the dilaton gradient in the throat limit from $-1/2Q$
to $-(1+\epsilon)/2Q$, where $\epsilon = (2b-1) \hat{\alpha}$.

After the coordinate transformation, we get
\begin{eqnarray}
d^{2}s_{string} & = &
        - \left( 1 - \frac{\hat{\alpha}}{2}x^{4}\right) dt^{2}
	+ \left(1-\frac{Q}{r}\right)^{2} f_{1}(r) dr^{2}
        + r^{2} d\Omega
	\nonumber \\
e^{-2\phi} & = & \left( 1-\frac{Q}{r} \right)^{1+\epsilon} f_{4}(r)
	\label{solution_string}
\end{eqnarray}
in the string metric, and
\begin{eqnarray}
d^{2}s & = &
        - \left(1-\frac{Q}{r}\right)^{1+\epsilon} f_{2}(r) dt^{2}
\nonumber \\
       &   & \mbox{} +
	\left(1-\frac{Q}{r}\right)^{-1+\epsilon} f_{3}(r) dr^{2}
        + r^{2} \left(1-\frac{Q}{r}\right)^{1+\epsilon} f_{4}(r) d\Omega
\label{eq:solution_Einstein}
\end{eqnarray}
in the Einstein metric,
where
\begin{eqnarray}
f_{1}(r) & = & 1 - \frac{\hat{\alpha}}{20} x(14x^{3}+9x^{2}+5x+2)
	\nonumber \\
f_{2}(r) & = & 1 - \frac{\hat{\alpha}}{40} x (11x^{3}+7x^{2}+16x+38)
			+ g(r)
	\nonumber \\
f_{3}(r) & = & 1 - \frac{\hat{\alpha}}{40} x (19x^{3}+25x^{2}+26x+42) 			+ g(r)
	\nonumber \\
f_{4}(r) & = & 1 - \frac{\hat{\alpha}}{40} x(-9x^{3}+7x^{2}+16x+38)
			+ g(r)
	\nonumber \\
g(r) & = & \frac{\hat{\alpha}}{60} b\,x(15x^{3}+32x^{2}+57x+120).
\end{eqnarray}
The solution describes an extremal black hole
whose singularity and horizon coincide.
However, the detailed form of the solution is not important
because $O(1/r^{3})$ terms do not have invariant meaning.
What are important to us are invariant relations like $M=M(Q)$.
Using (\ref{eq:mass}), we get that
the gravitational and inertial masses are the same and given by
\begin{equation}
M = \frac{Q}{2} - \frac{\alpha'}{40Q}.
\end{equation}
This is our main result.

It is not hard to see why $M(Q)$ is modified by higher dimensional
operators.
As is well-known, the throat of the extremal GHS black hole results
from the balance of curvature against the monopole magnetic field.
However, by including $O(\alpha')$ terms,
the curvature squared terms, {\it e.g.\/}, $R^{\mu \nu}R_{\mu \nu}$
cancel part of the monopole term $F_{\mu\nu}F^{\mu\nu}$.
Thus, for the balance to work at this order,
the black hole mass has to be slightly lighter than
the leading value for a given charge $Q$.

An interesting possibility arises by letting $Q$ get small.
Obviously, our perturbation breaks down for small $Q$,
but if the result were still valid, the solution might suggest
violation of the positive-energy theorem \cite{wald}.

\vspace{.1in}

\begin{center}
    {\Large {\bf Acknowledgements} }
\end{center}
\vspace{.1in}

I am grateful to J. Polchinski for his continuous help
and interest throughout the work.
I am also pleased to thank J. LaChapelle and W. Nelson for their comments on
the draft,
and thank S. Chaudhuri, G. Horowitz, T. Jacobson, and G. Kang
for useful conversation.
This research was supported in part by NSF Grant PHY 8904035 and 9116964.

\pagebreak

\appendix

\section{Field Redefinition Result}

Under the field redefinitions (\ref{eq:field_redef1}),
the coefficients of $O(\alpha')$ change as follows:
\begin{eqnarray*}
a'_{2} & = & a_{2}
	\\
a'_{3} & = & a_{3} - g_{1}
	\\
a'_{4} & = & a_{4} + \frac{1}{2}\,g_{1} + \frac{1}{2}(D-2)g_{4} - 2d_{1}
	\\
a'_{5} & = & a_{5} - 4g_{1} - g_{2}
	\\
a'_{6} & = & a_{6} + \frac{1}{2}\,g_{2} - 2D\,g_{4}
	+ \frac{1}{2}(D-2)g_{6} + 8d_{1} - 2d_{3}
	\\
a'_{7} & = & a_{7} + g_{1} + 2(D-1)g_{4} + \frac{1}{2}(D-2)g_{5}
	- 8d_{1} - 2d_{2}
	\\
a'_{8} & = & a_{8} + 2(D-1)g_{5} - 8d_{2}
	\\
a'_{9} & = & a_{9} + 3g_{2} - 2D\,g_{5} + 2(D-1)g_{6} + 8d_{2} - 8d_{3}
	\\
a'_{10} & = & a_{10} - 4g_{2} - 2D\,g_{6} + 8d_{3}
	\\
a'_{13} & = & a_{13} + \frac{1}{4}\,g_{3} - \frac{1}{4}(D-4)g_{7} + d_{4}
	\\
a'_{14} & = & a_{14}
	\\
a'_{15} & = & a_{15} + g_{1} - g_{3}
	\\
a'_{16} & = & a_{16} - \frac{1}{4}\,g_{1} + \frac{1}{2}\,g_{3}
	- \frac{1}{4}(D-4)g_{4} + \frac{1}{2}(D-2)g_{7} + d_{1} - 2d_{4}
	\\
a'_{17} & = & a_{17} - \frac{1}{4}\,g_{2} - g_{3} - \frac{1}{4}(D-4)g_{6}
	- 2D\,g_{7} + d_{3} + 8d_{4}
	\\
a'_{18} & = & a_{18} + \frac{3}{2}\,g_{3} - \frac{1}{4}(D-4)g_{5}
	+ 2(D-1)g_{7} + d_{2} - 8d_{4},
\end{eqnarray*}
where $D$ is the dimension of spacetime.
Not all the variations $\delta a_{i} = a'_{i} - a_{i}$
are independent of each other because
\begin{equation}
16\delta a_{4}-4\delta a_{6}-8\delta a_{7}+4\delta a_{8}
	+2\delta a_{9}+\delta a_{10} = 0.
\label{eq:invariant1}
\end{equation}
\begin{equation}
16\delta a_{3}-\frac{3}{2}\delta a_{5}+4\delta a_{8}
	+3\delta a_{9} +\frac{9}{4}\delta a_{10}+16\delta a_{13}
	+10\delta a_{15} +6\delta a_{17}+8\delta a_{18} = 0
\label{eq:invariant2}
\end{equation}

The actin (\ref{eq:full_action}) is transformed into (\ref{eq:GB_scheme})
by the following field redefinitions (we set $d=4$):
\begin{eqnarray*}
g_{1} & = & \frac{1}{2}+a_{3}
	\\
g_{2} & = & -2-4 a_{3}+a_{5}
	\\
g_{3} & = & \frac{1}{2}+a_{3}+a_{15}
	\\
g_{4} & = & \frac{1}{16}(3+8 a_{3}+8 a_{4}+6 a_{6}
		+4 a_{7}-2 a_{8}-a_{9}-4 d_{3})
	\\
g_{5} & = & \frac{3}{4}-6 a_{4}+\frac{3}{2}\,a_{6}
		+3 a_{7}-a_{8}-\frac{1}{4}\,a_{9}-d_{3}
	\\
g_{6} & = & \frac{5}{4}+2 a_{3}-2 a_{4}-\frac{1}{2}\,a_{5}
		+\frac{1}{2}\,a_{6}+a_{7}-\frac{1}{2}\,a_{8}
		-\frac{1}{4}\,a_{9}+d_{3}
	\\
g_{7} & = & \frac{1}{32}(-3+24 a_{3}+120 a_{4}-6 a_{6}
-28 a_{7}+6 a_{8}+a_{9}
	\nonumber \\
	& & \mbox{} \hspace{1.9in} +8 a_{15}+64 a_{16}-16 a_{18}
		+4 d_{3})
	\\
d_{1} & = & \frac{1}{32}(5+16 a_{3}+24 a_{4}+6 a_{6}+4 a_{7}
		-2 a_{8}-a_{9}-4d_{3})
	\\
d_{2} & = & \frac{1}{16}(9-72 a_{4}+18 a_{6}+36 a_{7}-10 a_{8}
		-3 a_{9}-12 d_{3})
	\\
d_{4} & = & \frac{1}{32}(3+24 a_{3}+72 a_{4}-12 a_{7}+2 a_{8}
		+12 a_{15}+48 a_{16}-8 a_{18}).
\end{eqnarray*}
After the redefinitions, the coefficients of $b\,(= a'_{13})$ and $c\,(=
a'_{18})$ are given by
\begin{eqnarray}
b & = & \frac{7}{32}+a_{3}+\frac{9}{4}\,a_{4}-\frac{3}{8}\,a_{7}
	+\frac{1}{16}\,a_{8}+a_{13}+\frac{5}{8}\,a_{15}
	+\frac{3}{2}\,a_{16}-\frac{1}{4}\,a_{18}
	\nonumber \\
c & = & \frac{3}{2}	-12 a_{4}-\frac{1}{4}a_{5}+\frac{3}{2}\,a_{6}
	+4 a_{7}-a_{8}-\frac{1}{4}\,a_{9}-4 a_{16}+a_{17}+2 a_{18}.
\label{eq:bc}
\end{eqnarray}

\pagebreak

\end{document}